\documentstyle[12pt,epsf]{article}
\newlength{\pubnumber} 

\catcode`\@=11
\@addtoreset{equation}{section}

\def\section{\@startsection{section}{1}{\z@}{3.5ex plus 1ex minus .2ex}
 {2.3ex plus .2ex}{\large\bf}}
\def\subsection{\@startsection{subsection}{2}{\z@}{2.3ex plus .2ex}
 {2.3ex plus .2ex}{\bf}}

\newfont{\mbm}{msbm10 scaled\magstep1}
\def\bb#1{\hbox{\mbm #1}}

\DeclareMathAlphabet\Scr{U}{rsf}{m}{n}

\begin{document}

\begin{titlepage}
\samepage{
\setcounter{page}{1}
\rightline{OUTP--02--25P}
\rightline{\tt hep-th/0206165}
\rightline{June 2002}
\vfill
\begin{center}
 {\Large \bf Partition functions of \\NAHE--based free fermionic
string models\\}
\vfill
\vfill {\large
	 Alon E. Faraggi\footnote{faraggi@thphys.ox.ac.uk}}\\
\vspace{.12in}
{\it  Theoretical Physics Department, University of Oxford,\\
Oxford OX1 3NP, United Kingdom\\}
\vspace{.025in}
\end{center}
\vfill
\begin{abstract}

The heterotic string free fermionic formulation produced 
a large class of three generation models, with an underlying SO(10)
GUT symmetry which is broken directly at the string level by Wilson lines.
A common subset of boundary condition basis vectors in these models
is the NAHE set, which corresponds to $\bb{Z}_2\times \bb{Z}_2$ orbifold
of an SO(12) Narain lattice, with $(h_{1,1},h_{2,1})=(27,3)$.
Alternatively, a manifold with the same data is obtained
by starting with a $\bb{Z}_2\times \bb{Z}_2$ orbifold at a
generic point on the
lattice, with $(h_{1,1},h_{2,1})=(51,3)$, and adding a freely acting
$\bb{Z}_2$ involution. The equivalence of the two constructions is proven
by examining the relevant partition functions. The 
explicit realization of the shift that reproduces the compactification
at the free fermionic point is found. It is shown that other closely related
shifts reproduce the same massless spectrum, but different massive
spectrum, thus demonstrating the utility of extracting information from
the full partition function. A freely acting involution of the type
discussed here, enables the use of Wilson lines to break the GUT symmetry
and can be utilized in non-perturbative studies of the free fermionic 
models.  

\end{abstract}
\smallskip}
\end{titlepage}

\setcounter{footnote}{0}

% ========================= DEFINITIONS ===================================
\def\beq{\begin{equation}}
\def\eeq{\end{equation}}
\def\beqn{\begin{eqnarray}}
\def\eeqn{\end{eqnarray}}

\def\no{\noindent }
\def\nolabel{\nonumber }
\def\ie{{\it i.e.}}
\def\eg{{\it e.g.}}
\def\half{{\textstyle{1\over 2}}}
\def\third{{\textstyle {1\over3}}}
\def\quarter{{\textstyle {1\over4}}}
\def\sixth{{\textstyle {1\over6}}}
\def\m{{\tt -}}
\def\p{{\tt +}}

\def\Tr{{\rm Tr}\, }
\def\tr{{\rm tr}\, }

\def\slash#1{#1\hskip-6pt/\hskip6pt}
\def\slk{\slash{k}}
\def\GeV{\,{\rm GeV}}
\def\TeV{\,{\rm TeV}}
\def\y{\,{\rm y}}
\def\SM{Standard--Model }
\def\SUSY{supersymmetry }
\def\SSSM{supersymmetric standard model}
\def\vev#1{\left\langle #1\right\rangle}
\def\l{\langle}
\def\r{\rangle}
\def\o#1{\frac{1}{#1}}

\def\Htw{{\tilde H}}
\def\chibar{{\overline{\chi}}}
\def\qbar{{\overline{q}}}
\def\ibar{{\overline{\imath}}}
\def\jbar{{\overline{\jmath}}}
\def\Hbar{{\overline{H}}}
\def\Qbar{{\overline{Q}}}
\def\abar{{\overline{a}}}
\def\alphabar{{\overline{\alpha}}}
\def\betabar{{\overline{\beta}}}
\def\tautwo{{ \tau_2 }}
\def\thetatwo{{ \vartheta_2 }}
\def\thetathree{{ \vartheta_3 }}
\def\thetafour{{ \vartheta_4 }}
\def\ttwo{{\vartheta_2}}
\def\tthree{{\vartheta_3}}
\def\tfour{{\vartheta_4}}
\def\ti{{\vartheta_i}}
\def\tj{{\vartheta_j}}
\def\tk{{\vartheta_k}}
\def\calF{{\cal F}}
\def\smallmatrix#1#2#3#4{{ {{#1}~{#2}\choose{#3}~{#4}} }}
\def\ab{{\alpha\beta}}
\def\Minv{{ (M^{-1}_\ab)_{ij} }}
\def\bone{{\bf 1}}
\def\ii{{(i)}}
\def\V{{\bf V}}
\def\N{{\bf N}}

% for basis vectors:
\def\b{{\bf b}}
\def\S{{\bf S}}
\def\X{{\bf X}}
\def\I{{\bf I}}
\def\mb{{\mathbf b}}
\def\mS{{\mathbf S}}
\def\mX{{\mathbf X}}
\def\mI{{\mathbf I}}
\def\balpha{{\mathbf \alpha}}
\def\bbeta{{\mathbf \beta}}
\def\bgamma{{\mathbf \gamma}}
\def\bxi{{\mathbf \xi}}

\def\t#1#2{{ \Theta\left\lbrack \matrix{ {#1}\cr {#2}\cr }\right\rbrack }}
\def\C#1#2{{ C\left\lbrack \matrix{ {#1}\cr {#2}\cr }\right\rbrack }}
\def\tp#1#2{{ \Theta'\left\lbrack \matrix{ {#1}\cr {#2}\cr }\right\rbrack }}
\def\tpp#1#2{{ \Theta''\left\lbrack \matrix{ {#1}\cr {#2}\cr }\right\rbrack }}
\def\l{\langle}
\def\r{\rangle}

%================== BLACKBOARD BOLD CHARACTERS ==============================

\def\inbar{\,\vrule height1.5ex width.4pt depth0pt}

\def\IC{\relax\hbox{$\inbar\kern-.3em{\rm C}$}}
\def\IQ{\relax\hbox{$\inbar\kern-.3em{\rm Q}$}}
\def\IR{\relax{\rm I\kern-.18em R}}
 \font\cmss=cmss10 \font\cmsss=cmss10 at 7pt
\def\IZ{\relax\ifmmode\mathchoice
 {\hbox{\cmss Z\kern-.4em Z}}{\hbox{\cmss Z\kern-.4em Z}}
 {\lower.9pt\hbox{\cmsss Z\kern-.4em Z}}
 {\lower1.2pt\hbox{\cmsss Z\kern-.4em Z}}\else{\cmss Z\kern-.4em Z}\fi}

%========================================================================
%          MACROS FOR REFERENCES
%========================================================================
\def\AEF{A.E. Faraggi}
\def\NPB#1#2#3{{\it Nucl.\ Phys.}\/ {\bf B#1} (#2) #3}
\def\PLB#1#2#3{{\it Phys.\ Lett.}\/ {\bf B#1} (#2) #3}
\def\PRD#1#2#3{{\it Phys.\ Rev.}\/ {\bf D#1} (#2) #3}
\def\PRL#1#2#3{{\it Phys.\ Rev.\ Lett.}\/ {\bf #1} (#2) #3}
\def\PRT#1#2#3{{\it Phys.\ Rep.}\/ {\bf#1} (#2) #3}
\def\MODA#1#2#3{{\it Mod.\ Phys.\ Lett.}\/ {\bf A#1} (#2) #3}
\def\IJMP#1#2#3{{\it Int.\ J.\ Mod.\ Phys.}\/ {\bf A#1} (#2) #3}
\def\nuvc#1#2#3{{\it Nuovo Cimento}\/ {\bf #1A} (#2) #3}
\def\RPP#1#2#3{{\it Rept.\ Prog.\ Phys.}\/ {\bf #1} (#2) #3}
\def\etal{{\it et al\/}}

%==============================================================================
\hyphenation{su-per-sym-met-ric non-su-per-sym-met-ric}
\hyphenation{space-time-super-sym-met-ric}
\hyphenation{mod-u-lar mod-u-lar--in-var-i-ant}
%==============================================================================

%============================== SECTION 1 ============================

\setcounter{footnote}{0}
\section{Introduction}
Grand unification, and its incarnation in the form of heterotic-string
unification \cite{heterotic},
is the only extension of the Standard Model that is
rooted in the structure of the Standard Model itself. In this context
the most realistic string models discovered to date have been constructed 
in the heterotic string free fermionic formulation. While this may be an
accident, it may also reflect on deeper, yet undiscovered, properties
of string theory. It is therefore imperative to enhance our understanding
of this particular class of models, with the hope that it will shed further
light on their properties, and possibly yield deeper insight
into the dynamics of string theory.

An important feature of the realistic free fermionic
models is their underlying $\bb{Z}_2\times \bb{Z}_2$
orbifold structure. Many of the encouraging phenomenological
characteristics of the realistic free fermionic
models are rooted in this
structure, including the three generations arising from the 
three twisted sectors, and the canonical SO(10) embedding for
the weak hyper-charge. To see more precisely this
orbifold correspondence, recall that 
the free fermionic models are generated by a set of basis vectors
which define the transformation properties of the world-sheet
fermions as they are transported around loops on the string world sheet.
A large set of realistic free fermionic models contains
a subset of boundary conditions, the so-called NAHE set, which
can be seen
to correspond to $\bb{Z}_2\times \bb{Z}_2$ orbifold compactification
with the standard embedding of the gauge connection~\cite{foc}.
This underlying free fermionic model contains 24 generations from the
twisted sectors, as well as three additional 
generation/anti-generation pairs from the untwisted sector.
At the free fermionic point in the Narain moduli space~\cite{Narain},
both the metric and the antisymmetric background fields
are non-trivial, leading to an SO(12) enhanced symmetry group.
The action of the $\bb{Z}_2\times \bb{Z}_2$ twisting on the
SO(12) Narain lattice then gives rise to a model
with $(h_{11},h_{21})=(27,3)$, matching the data of
the free fermionic model. It is noted that this data differs
from that of the $\bb{Z}_2\times \bb{Z}_2$ orbifold at a generic point in the
moduli space of $(T_2)^3$, which yields $(h_{11},h_{21})=(51,3)$. 
We refer to the (51,3) and (27,3) $\bb{Z}_2\times \bb{Z}_2$ orbifold models
as ${\Scr X}_1$ and ${\Scr X}_2$ respectively. 

While the free fermionic construction provides most useful tools to
analyze the spectrum and superpotential interactions in a given string
model, its drawback is that it is formulated at a fixed point in the
compactification moduli space. The moduli dependence of physical quantities
may therefore only be studied by including world-sheet Thirring
interactions, which may be cumbersome. On the other hand, 
the moduli dependence is more readily extracted by constructing the 
models in a geometrical, or orbifold \cite{dhvw}, formalism.
Another important advantage of the geometrical approach
is that it might provide a closer link to the various strong-weak
coupling dualities that have been unraveled in the past few years.
In the context of the realistic free fermionic models preliminary 
investigations have been attempted by relating the F-theory
compactification on $X_2$ to the studies of F-theory compactification
on $X_1$. This study highlighted the potential
relevance of Calabi-Yau manifolds, which possess a bi-section
but not a global section and which was further investigated in
\cite{bisection}. 

The key ingredient in studying the F-theory compactification on 
the free fermionic $\bb{Z}_2\times \bb{Z}_2$ orbifold was to connect it to the 
$X_1$ by a freely acting twist or shift
\cite{befnq}. Under the freely acting shift, pairs of
twisted sector fixed points are identified. Hence reducing the 
total number of fixed points from 48 to 24. It is noted that at
the level of the toroidal compactification, the ${\rm SO}(4)^3$
and SO(12) lattices are continuously connected by 
varying the parameters of the background metric and antisymmetric 
tensor. However, this cannot be done while preserving the
$\bb{Z}_2\times \bb{Z}_2$ invariance, because 
the continuous interpolation cannot change the Euler characteristic.
Indeed, part of the geometric moduli are projected out by the orbifold
action and, as a result, though the two toroidal models are in the same
moduli space, the two $\bb{Z}_2\times \bb{Z}_2$ orbifold models are not. 

The connection of these studies to the free fermionic $\bb{Z}_2\times \bb{Z}_2$
orbifold model therefore rests on the conjecture that the model
constructed with the additional freely acting shift is identical 
to the model at the free fermionic point in the Narain moduli
space, i.e. to the $\bb{Z}_2\times \bb{Z}_2$ orbifold on the SO(12) lattice.
However, the validity of this conjecture is far from obvious. 
While the massless spectrum and symmetries of the two models
match, their massive spectrum may differ.  

In this paper we therefore undertake the task of proving this 
conjecture. This is achieved by constructing the partition
function of the $\bb{Z}_2\times \bb{Z}_2$ orbifold on an SO(12) lattice, 
and the partition function of the $\bb{Z}_2\times \bb{Z}_2$ orbifold
on a generic $(T_2)^3$ lattice. We then show that adding the 
freely acting shift to the latter and fixing the radii of the
compactified dimensions at the self-dual point reproduces
the partition function on the SO(12) lattice, hence proving that
the models are identical. However, we show that this
matching is highly non-trivial and is obtained only for a specific
form of the freely acting shifts which affects simultaneously 
momenta and windings. In contrast freely acting shifts that act
only on momenta or winding do not reproduce the partition
function of the free fermionic model. Thus, while the spectra 
of the three models match at the massless level, they in 
general differ at the massive level. Hence demonstrating
the usefulness of gaining additional valuable information from
the construction of the partition function. 
Additionally we discuss in this paper the general structure 
of the partition functions of NAHE based free fermionic models.

\setcounter{footnote}{0}
\section{Realistic free fermionic models - general structure}\label{gs}
In this section we recapitulate the main structure of
the realistic free fermionic models.
The notation 
and details of the construction of these
models are given elsewhere \cite{fsu5,fny,alr,euslm,nahe,cfn,cfs}.
In the free fermionic formulation of the heterotic string
in four dimensions all the world-sheet
degrees of freedom  required to cancel
the conformal anomaly are represented in terms of free fermions
propagating on the string world-sheet \cite{fff}.
In the light-cone gauge the world-sheet field content consists
of two transverse left- and right-moving space-time coordinate bosons,
$X_{1,2}^\mu$ and ${\bar X}_{1,2}^\mu$,
and their left-moving fermionic superpartners $\psi^\mu_{1,2}$,
and additional 62 purely internal
Majorana-Weyl fermions, of which 18 are left-moving,
$\chi^{I}$, and 44 are right-moving, $\phi^a$.
In the supersymmetric sector the world-sheet supersymmetry is realized
non-linearly and the world-sheet supercurrent is given by
$T_F=\psi^\mu\partial X_\mu+i\chi^Iy^I\omega^I,~(I=1,\cdots,6).$
The $\{\chi^{I},y^I,\omega^I\}~(I=1,\cdots,6)$ are 18 real free
fermions transforming as the adjoint representation of ${\rm SU}(2)^6$.
Under parallel transport around a non-contractible loop on the toroidal
world-sheet the fermionic fields pick up a phase,
$
f~\rightarrow~-{\rm e}^{i\pi\alpha(f)}f~,~~\alpha(f)\in(-1,+1].
$
Each set of specified
phases for all world-sheet fermions, around all the non-contractible
loops is called the spin structure of the model. Such spin structures
are usually given is the form of 64 dimensional boundary condition vectors,
with each element of the vector specifying the phase of the corresponding
world-sheet fermion. The basis vectors are constrained by string consistency
requirements and completely determine the vacuum structure of the model.
The physical spectrum is obtained by applying the generalized GSO projections.

The boundary condition basis defining a typical 
realistic free fermionic heterotic string model is 
constructed in two stages. 
The first stage consists of the NAHE set,
which is a set of five boundary condition basis vectors, 
$\{ 1 ,S,b_1,b_2,b_3\}$ \cite{nahe}. 
The gauge group after imposing the GSO projections induced
by the NAHE set is ${\rm SO} (10)\times {\rm SO}(6)^3\times {\rm E}_8$
with ${\Scr N}=1$ supersymmetry. The space-time vector bosons that generate
the gauge group arise from the Neveu-Schwarz sector and
from the sector $\xi_2\equiv 1+b_1+b_2+b_3$. The Neveu-Schwarz sector
produces the generators of ${\rm SO}(10)\times {\rm SO}(6)^3\times 
{\rm SO}(16)$. The $\xi_2$-sector produces the spinorial 128
of SO(16) and completes the hidden gauge group to ${\rm E}_8$.
The NAHE set divides the internal world-sheet 
fermions in the following way: ${\bar\phi}^{1,\cdots,8}$ generate the 
hidden ${\rm E}_8$ gauge group, ${\bar\psi}^{1,\cdots,5}$ generate the SO(10) 
gauge group, and $\{{\bar y}^{3,\cdots,6},{\bar\eta}^1\}$,  
$\{{\bar y}^1,{\bar y}^2,{\bar\omega}^5,{\bar\omega}^6,{\bar\eta}^2\}$,
$\{{\bar\omega}^{1,\cdots,4},{\bar\eta}^3\}$ generate the three horizontal 
${\rm SO}(6)^3$ symmetries. The left-moving $\{y,\omega\}$ states are divided 
into $\{{y}^{3,\cdots,6}\}$,  
$\{{y}^1,{y}^2,{\omega}^5,{\omega}^6\}$,
$\{{\omega}^{1,\cdots,4}\}$ and $\chi^{12}$, $\chi^{34}$, $\chi^{56}$ 
generate the left-moving ${\Scr N}=2$ world-sheet supersymmetry.
At the level of the NAHE set the sectors $b_1$, $b_2$ and $b_3$
produce 48 multiplets, 16 from each, in the $16$ 
representation of SO(10). The states from the sectors $b_j$
are singlets of the hidden ${\rm E}_8$ gauge group and transform 
under the horizontal ${\rm SO}(6)_j$ $(j=1,2,3)$ symmetries. This structure
is common to all the realistic free fermionic models.

The second stage of the
construction consists of adding to the 
NAHE set three (or four) additional boundary condition basis vectors,
typically denoted by $\{\alpha,\beta,\gamma\}$. 
These additional basis vectors reduce the number of generations
to three chiral generations, one from each of the sectors $b_1$,
$b_2$ and $b_3$, and simultaneously break the four dimensional
gauge group. The assignment of boundary conditions to
$\{{\bar\psi}^{1,\cdots,5}\}$ breaks SO(10) to one of its subgroups
${\rm SU}(5)\times {\rm U}(1)$ \cite{fsu5}, ${\rm SO}(6)\times {\rm SO}(4)$ 
\cite{alr},
${\rm SU}(3)\times {\rm SU}(2)\times {\rm U}(1)^2$ \cite{fny,euslm,cfn}
or ${\rm SU}(3)\times {\rm SU}(2)^2\times {\rm U}(1)$ \cite{cfs}.
Similarly, the hidden ${\rm E}_8$ symmetry is broken to one of its
subgroups by the basis vectors which extend the NAHE set.
The flavour ${\rm SO}(6)^3$ symmetries in the NAHE-based models
are always broken to flavour U(1) symmetries, as the breaking
of these symmetries is correlated with the number of chiral
generations. Three such ${\rm U}(1)_j$ symmetries are always obtained
in the NAHE based free fermionic models, from the subgroup
of the observable ${\rm E}_8$, which is orthogonal to SO(10).
These are produced by the world-sheet currents ${\bar\eta}{\bar\eta}^*$
($j=1,2,3$), which are part of the Cartan sub-algebra of the
observable ${\rm E}_8$. Additional unbroken U(1) symmetries, denoted
typically by ${\rm U}(1)_j$ ($j=4,5,...$), arise by pairing two real
fermions from the sets $\{{\bar y}^{3,\cdots,6}\}$,
$\{{\bar y}^{1,2},{\bar\omega}^{5,6}\}$ and
$\{{\bar\omega}^{1,\cdots,4}\}$. The final observable gauge
group depends on the number of such pairings. 

The correspondence of the NAHE-based free fermionic models
with the orbifold construction is illustrated
by extending the NAHE set, $\{ 1,S,b_1,b_2,b_3\}$, by one additional
boundary condition basis vector \cite{foc}
\beq
\xi_1=(0,\cdots,0\vert{\underbrace{1,\cdots,1}_{{\bar\psi^{1,\cdots,5}},
{\bar\eta^{1,2,3}}}},0,\cdots,0)~.
\label{vectorx}
\eeq
With a suitable choice of the GSO projection coefficients the 
model possesses an ${\rm SO}(4)^3\times {\rm E}_6\times {\rm U}(1)^2
\times {\rm E}_8$ gauge group
and ${\Scr N}=1$ space-time supersymmetry. The matter fields
include 24 generations in the 27 representation of
${\rm E}_6$, eight from each of the sectors $b_1\oplus b_1+\xi_1$,
$b_2\oplus b_2+\xi_1$ and $b_3\oplus b_3+\xi_1$.
Three additional 27 and $\overline{27}$ pairs are obtained
from the Neveu-Schwarz $\oplus~\xi_1$ sector.

To construct the model in the orbifold formulation one starts
with the compactification on a torus with nontrivial background
fields \cite{Narain}.
The subset of basis vectors
\beq
\{ 1,S,\xi_1,\xi_2\}
\label{neq4set}
\eeq
generates a toroidally-compactified model with ${\Scr N}=4$ space-time
supersymmetry and ${\rm SO}(12)\times {\rm E}_8\times {\rm E}_8$ gauge group.
The same model is obtained in the geometric (bosonic) language
by tuning the background fields to the values corresponding to
the SO(12) lattice. The 
metric of the six-dimensional compactified
manifold is then the Cartan matrix of SO(12), 
while the antisymmetric tensor is given by
\beq
B_{ij}=\cases{
G_{ij}&;\ $i>j$,\cr
0&;\ $i=j$,\cr
-G_{ij}&;\ $i<j$.\cr}
\label{bso12}
\eeq
When all the radii of the six-dimensional compactified
manifold are fixed at $R_I=\sqrt2$, it is seen that the
left- and right-moving momenta
$%\beq
P^I_{R,L}=[m_i-{1\over2}(B_{ij}{\pm}G_{ij})n_j]{e_i^I}^*
%\label{lrmomenta}
$%\eeq
reproduce the massless root vectors in the lattice of
SO(12). Here $e^i=\{e_i^I\}$ are six linearly-independent
vielbeins normalised so that $(e_i)^2=2$.
The ${e_i^I}^*$ are dual to the $e_i$, with
$e_i^*\cdot e_j=\delta_{ij}$.

Adding the two basis vectors $b_1$ and $b_2$ to the set
(\ref{neq4set}) corresponds to the $\bb{Z}_2\times \bb{Z}_2$
orbifold model with standard embedding.
Starting from the Narain model with ${\rm SO}(12)\times 
{\rm E}_8\times {\rm E}_8$
symmetry~\cite{Narain}, and applying the $\bb{Z}_2\times \bb{Z}_2$ 
twist on the
internal coordinates, reproduces
the spectrum of the free-fermion model
with the six-dimensional basis set
$\{ 1,S,\xi_1,\xi_2,b_1,b_2\}$.
The Euler characteristic of this model is 48 with $h_{11}=27$ and
$h_{21}=3$.

It is noted that the effect of the additional basis vector $\xi_1$ of eq.
(\ref{vectorx}), is to separate the gauge degrees of freedom, spanned by
the world-sheet fermions $\{{\bar\psi}^{1,\cdots,5},
{\bar\eta}^{1},{\bar\eta}^{2},{\bar\eta}^{3},{\bar\phi}^{1,\cdots,8}\}$,
from the internal compactified degrees of freedom $\{y,\omega\vert
{\bar y},{\bar\omega}\}^{1,\cdots,6}$. 
In the realistic free fermionic
models this is achieved by the vector $2\gamma$ \cite{foc}, with
\beq
2\gamma=(0,\cdots,0\vert{\underbrace{1,\cdots,1}_{{\bar\psi^{1,\cdots,5}},
{\bar\eta^{1,2,3}} {\bar\phi}^{1,\cdots,4}} },0,\cdots,0)~,
\label{vector2gamma}
\eeq
which breaks the ${\rm E}_8\times {\rm E}_8$ symmetry to ${\rm SO}(16)\times 
{\rm SO}(16)$. 
The $\bb{Z}_2\times \bb{Z}_2$ twist breaks the gauge symmetry to
${\rm SO}(4)^3\times {\rm SO}(10)\times {\rm U}(1)^3\times {\rm SO}(16)$.
The orbifold still yields a model with 24 generations,
eight from each twisted sector,
but now the generations are in the chiral 16 representation
of SO(10), rather than in the 27 of ${\rm E}_6$. The same model can
be realized with the set
$\{ 1,S,\xi_1,\xi_2,b_1,b_2\}$,
by projecting out the $16\oplus{\overline{16}}$
from the $\xi_1$-sector taking
\beq
c{\xi_1\choose \xi_2}\rightarrow -c{\xi_1\choose \xi_2}.
\label{changec}
\eeq
This choice also projects out the massless vector bosons in the
128 of SO(16) in the hidden-sector ${\rm E}_8$ gauge group, thereby
breaking the ${\rm E}_6\times {\rm E}_8$ symmetry to
${\rm SO}(10)\times {\rm U}(1)\times {\rm SO}(16)$.
The freedom in ({\ref{changec}) corresponds to 
a discrete torsion in the toroidal orbifold model.
At the level of the ${\Scr N}=4$ 
Narain model generated by the set (\ref{neq4set}),
we can define two models, ${\Scr Z}_+$ and ${\Scr Z}_-$, depending on the sign
of the discrete torsion in eq. (\ref{changec}). The first, say ${\Scr Z}_+$,
produces the ${\rm E}_8\times {\rm E}_8$ model, whereas the second, say 
${\Scr Z}_-$, produces the ${\rm SO}(16)\times {\rm SO}(16)$ model. 
However, the $\bb{Z}_2\times 
\bb{Z}_2$
twist acts identically in the two models, and their physical characteristics
differ only due to the discrete torsion eq. (\ref{changec}). 
The important aspect, however, is the separation, by the extended NAHE
set, of the world-sheet fermionic degrees of freedom corresponding 
to the space-time gauge degrees of freedom, from those 
corresponding to the internal compactified dimensions.

This analysis confirms that the $\bb{Z}_2\times \bb{Z}_2$ orbifold on the
SO(12) Narain lattice is indeed at the core of the
realistic free fermionic models. However, it
differs from the $\bb{Z}_2\times \bb{Z}_2$ orbifold on 
$T_2^1\times T_2^2\times T_2^3$ with $(h_{11},h_{21})=(51,3)$.
In \cite{befnq} it was shown that the two models are connected
by adding a freely acting twist or shift to the $X_1$ model. 
Let us first start with the compactified
$T^1_2\times T^2_2\times T^3_2$ torus parametrised by
three complex coordinates $z_1$, $z_2$ and $z_3$,
with the identification
\beq
z_i=z_i + 1\,, \qquad z_i=z_i+\tau_i \,,
\label{t2cube}
\eeq
where $\tau$ is the complex parameter of each
$T_2$ torus. 
With the identification $z_i\rightarrow-z_i$, a single torus 
has four fixed points at
\beq
z_i=\{0,{\textstyle{1\over 2}},{\textstyle{1\over 2}}\,\tau,
{\textstyle{1\over 2}} (1+\tau) \}.
\label{fixedtau}
\eeq
With the two $\bb{Z}_2$ twists 
\beqn
&& \alpha:(z_1,z_2,z_3)\rightarrow(-z_1,-z_2,~~z_3) \,,
\cr
&&  \beta:(z_1,z_2,z_3)\rightarrow(~~z_1,-z_2,-z_3)\,,
\label{alphabeta}
\eeqn
there are three twisted sectors in this model, $\alpha$,
$\beta$ and $\alpha\beta=\alpha\cdot\beta$, each producing
16 fixed tori, for a total of 48. Adding 
to the model generated by the $\bb{Z}_2\times \bb{Z}_2$
twist in (\ref{alphabeta}), the additional shift
\beq
\gamma:(z_1,z_2,z_3)\rightarrow(z_1+{\textstyle{1\over2}},z_2+
{\textstyle{1\over2}},z_3+{\textstyle{1\over2}})
\label{gammashift}
\eeq
produces again fixed tori from the three
twisted sectors $\alpha$, $\beta$ and $\alpha\beta$.
The product of the $\gamma$ shift in (\ref{gammashift})
with any of the twisted sectors does not produce any additional
fixed tori. Therefore, this shift acts freely.
Under the action of the $\gamma$-shift,
the fixed tori from each twisted sector are paired.
Therefore, $\gamma$ reduces
the total number of fixed tori from the twisted sectors
by a factor of ${1 \over 2}$, 
with $(h_{11},h_{21})=(27,3)$. This model therefore
reproduces the data of the $\bb{Z}_2\times \bb{Z}_2$ orbifold
at the free-fermion point in the Narain moduli space.

\setcounter{footnote}{0}
\section{NAHE-based partition functions}

In the previous section we showed that the freely
acting shift (\ref{gammashift}), added to the $\bb{Z}_2\times 
\bb{Z}_2$ orbifold
at a generic point of $T_2^1\times T_2^2\times T_2^3$
Eq. (\ref{alphabeta}) reproduces the data of the $\bb{Z}_2\times \bb{Z}_2$
orbifold acting on the SO(12) lattice, which coincides with the
lattice at the free fermionic point. However, this observation
does not prove that the vacuum which includes the shift
is identical to the free fermionic model. While the 
massless spectrum of the two models may coincide
their massive excitations, in general, may differ.
To examine the matching of the massive spectra
we construct the partition function of the $\bb{Z}_2\times \bb{Z}_2$
orbifold of an SO(12) lattice and subsequently
that of the model at a generic point including the
shift. In effect since the action of the $\bb{Z}_2\times \bb{Z}_2$
orbifold in the two cases is identical our problem
reduces to proving the existence of a freely
acting shift that reproduces the partition function of the
SO(12) lattice
at the free fermionic point. Then since the action of 
the shift and the orbifold projections are commuting
it follows that the two $\bb{Z}_2\times \bb{Z}_2$ orbifolds
are identical.

On the compact coordinates there are actually three inequivalent ways
in which the shifts
can act. In the more familiar case, they simply translate a generic point 
by half the
length of the circle. As usual, the presence of windings in string 
theory allows shifts on the T-dual circle, or even asymmetric ones, that 
act both on the circle and on its dual. More concretely, for a circle of
length $2 \pi R$, one can have the following options \cite{vwaaf}:
\beqn
A_1\;:&& X_{\rm L,R} \to X_{\rm L,R} + {\textstyle{1\over 2}} \pi R \,,
\nonumber \\
A_2\;:&& X_{\rm L,R} \to X_{\rm L,R} + {\textstyle{1\over 2}} \left(
\pi R \pm {\pi \alpha ' \over R} \right) \,, 
\nonumber \\
A_3\;:&& X_{\rm L,R} \to X_{\rm L,R} \pm {\textstyle{1\over 2}} {\pi \alpha'
\over R} \,.
\eeqn
There is, however, a crucial difference among these three choices: while
$A_1$ and $A_3$ shifts can act consistently on any number of coordinates,
level-matching requires instead that the $A_2$-shifts act on (mod) four real 
coordinates. 

We can now proceed to deform the free fermionic models to a (connected) 
generic point in moduli space. As we already noticed, the $\bb{Z}_2 
\times \bb{Z}_2$ orbifold and the shifts are commuting operations,
and thus it suffices to find the correct shift that would correspond
the the heterotic string on the SO(12) lattice with
discrete torsion 
\beqn
{\Scr Z}_-&=& (V_8-S_8)\left[\left( |O_{12}|^2+|V_{12}|^2 \right) 
\left(\bar O_{16} \bar O_{16}+ \bar C_{16} \bar C_{16}\right)\right.
\cr
&& + \left( |S_{12}|^2+|C_{12}|^2 \right) \left( \bar S_{16} \bar S_{16}+
\bar V_{16} \bar V_{16}\right)
\cr
&& + \left( O_{12} \bar V_{12} + V_{12} \bar O_{12} \right)
\left( \bar S_{16} \bar V_{16} + \bar V_{16} \bar S_{16}\right)
\cr
&& + \left. \left( S_{12} \bar C_{12} + C_{12} \bar S_{12} \right)
\left( \bar O_{16} \bar C_{16} + \bar C_{16} \bar O_{16} \right) \right] \,.
\label{zminus}
\eeqn
or without
\beq
{\Scr Z}_+=(V_8-S_8)\left[|O_{12}|^2+|V_{12}|^2+|S_{12}|^2+|C_{12}|^2\right]
\left( \bar O_{16} + \bar S_{16}\right) \left( \bar O_{16} + \bar S_{16}
\right) \,, \label{zplus}
\eeq
discrete torsion. Here we have written ${\Scr Z}_{\pm}$ in terms of level-one
${\rm SO} (2n)$ characters (see, for instance, \cite{as})
\beqn
O_{2n} &=& {\textstyle{1\over 2}} \left( {\vartheta_3^n \over \eta^n} +
{\vartheta_4^n \over \eta^n}\right) \,,
\nonumber \\
V_{2n} &=& {\textstyle{1\over 2}} \left( {\vartheta_3^n \over \eta^n} -
{\vartheta_4^n \over \eta^n}\right) \,,
\nonumber \\
S_{2n} &=& {\textstyle{1\over 2}} \left( {\vartheta_2^n \over \eta^n} +
i^{-n} {\vartheta_1^n \over \eta^n} \right) \,,
\nonumber \\
C_{2n} &=& {\textstyle{1\over 2}} \left( {\vartheta_2^n \over \eta^n} -
i^{-n} {\vartheta_1^n \over \eta^n} \right) \,.
\eeqn

These two models can actually be connected by the orbifold
\beq
{\Scr Z}_- = {\Scr Z}_+ / a \otimes b \,,
\eeq
with
\beqn
a &=& (-1)^{F_{\rm L}^{\rm int} + F_\xi^1} \,,
\nonumber \\
b &=& (-1)^{F_{\rm L}^{\rm int} + F_\xi^2} \,. \label{orbzpm}
\eeqn
Therefore, in the following discussion we shall focus on ${\Scr Z}_+$,
and the corresponding results for ${\Scr Z}_-$, cumbersome as they may be, 
can be obtained applying the orbifold projection (\ref{orbzpm}). 
Starting from $Z_+$ we can obtain 
the partition functions of NAHE--based free fermionic models
by applying the $\bb{Z}_2\times \bb{Z}_2$ orbifold projection 
(\ref{alphabeta}). As the result is somewhat lengthy
we do not display it here explicitly. 

Instead our problem is to find the shift that when acting on the 
lattice $T_2^1\otimes T_2^2\otimes T_2^3$ at a generic point in
the moduli space reproduces the $SO(12)$ lattice when the radii
are fixed at the self--dual point $R=\sqrt{\alpha^\prime}$.
Let us consider for simplicity the case of six orthogonal circles or 
radii $R_i$. The partition function reads
\beq
{\Scr Z}_+ = (V_8 - S_8) \, \left( \sum_{m,n} \Lambda_{m,n}
\right)^{\otimes 6}\, \left( \bar O _{16} + \bar S_{16} \right) \left(
\bar O _{16} + \bar S_{16} \right)\,,
\eeq
where as usual, for each circle,
\beq
p_{\rm L,R}^i = {m_i \over R_i} \pm {n_i R_i \over \alpha '} \,,
\eeq
and
\beq
\Lambda_{m,n} = {q^{{\alpha ' \over 4} 
p_{\rm L}^2} \, \bar q ^{{\alpha ' \over 4} p_{\rm R}^2} \over |\eta|^2}\,.
\eeq
We can now act with the $\bb{Z}_2\times \bb{Z}_2$ shifts generated by
\beqn
g\;: & & (A_2 , A_2 ,0 ) \,,
\nonumber \\
h\;: & & (0, A_2 , A_2 ) \,, \label{gfh}
\eeqn
where each $A_2$ acts on a complex coordinate. The resulting partition 
function then reads
\beqn
{\Scr Z}_+ &=& {\textstyle{1\over 4}}\, (V_8 - S_8) 
\sum_{m_i , n_i}  \left\{ \left[ 1 + (-1)^{m_1 + m_2 + m_3 + m_4 + n_1 + n_2 +
n_3 + n_4} \right. \right.
\nonumber \\
& & \left. + (-1)^{m_1 + m_2 + m_5 + m_6 + n_1 + n_2 +
n_5 + n_6} + (-1)^{m_3 + m_4 + m_5 + m_6 + n_3 + n_4 +
n_5 + n_6} \right]  
\nonumber \\
& & \left. \times \left( \Lambda_{m_i , n_i}^{1,\ldots ,6} 
+ \Lambda^{1,\ldots,4}_{m_i + {1\over 2}, n_i + {1\over 2}} 
\Lambda^{5,6}_{m_i , n_i} 
+ \Lambda^{1,2,5,6}_{m_i + {1\over 2}, n_i + {1\over 2}} 
\Lambda^{3,4}_{m_i , n_i}
+ \Lambda^{1,2}_{m_i , n_i} 
\Lambda^{3,4,5,6}_{m_i + {1\over 2}, n_i + {1\over 2}} 
\right) \right\}
\nonumber \\
& & \times
\left( \bar O _{16} + \bar S_{16} \right) \left( \bar O_{16} + \bar S_{16}
\right) \label{zpshift}
\eeqn

After some tedious algebra, it is then possible to show that, once evaluated
at the self-dual radius $R_i = \sqrt{\alpha '}$, the 
partition function (\ref{zpshift}) reproduces that at the SO(12) point
(\ref{zplus}). To this end, it suffices to notice that
\beqn
\sum_{m,n} \Lambda_{m,n} (R=\sqrt{\alpha '}) &=& |\chi_0 |^2 + |\chi_{1\over 2}
|^2 \,,
\nonumber \\
\sum_{m,n} (-1)^{m+n} \Lambda_{m,n} (R = \sqrt{\alpha '}) &=&
|\chi_0 |^2 - |\chi_{1\over 2} |^2 \,,
\nonumber \\
\sum_{m,n} \Lambda_{m + {1\over 2} , n + {1\over 2}} (R = \sqrt{\alpha '})
&=& \chi_0 \bar\chi_{1\over 2} + \chi_{1\over 2} \bar \chi_0 \,,
\nonumber \\
\sum_{m,n} (-1)^{m+n} \Lambda_{m + {1\over 2} , n + {1\over 2}} 
(R = \sqrt{\alpha '}) &=& \chi_{1\over 2} \bar \chi_0 -
\chi_0 \bar\chi_{1\over 2}  \,,
\eeqn
where
\beqn
\chi_0 &=& \sum_\ell q^{\ell^2} \,,
\nonumber \\
\chi_{1\over 2} &=& \sum_\ell q^{(\ell + {1\over 2})^2} \,,
\eeqn
are the two level-one SU(2) characters, while, standard branching rules,
decompose the SO(12) characters into products of SU(2) ones. For instance,
\beqn
O_{12} &=& \chi_0 \chi_0 \chi_0 \chi_0 \chi_0 \chi_0 +
\chi_0 \chi_0 \chi_{1\over 2} \chi_{1\over 2}
\chi_{1\over 2} \chi_{1\over 2}
\nonumber \\
& &\chi_{1\over 2} \chi_{1\over 2} \chi_0 
\chi_0 \chi_{1\over 2} \chi_{1\over 2} +
\chi_{1\over 2} \chi_{1\over 2} \chi_{1\over 2} \chi_{1\over 2} 
\chi_0 \chi_0 \,.
\eeqn

Let us now consider the shifts given in Eq. (\ref{gammashift}),
and similarly the analogous freely acting shift given by 
$(A_3,A_3,A_3)$. 
The $SO(4)$ lattice takes the form
\beq
\Lambda_{SO(4)}={1\over{\vert\eta\vert^4}}
\left(\vert\chi_0\vert^2+\vert\chi_{1\over2}\vert^2\right)
\left(\vert\chi_0\vert^2+\vert\chi_{1\over2}\vert^2\right)
\eeq
The effect of the free acting shift Eq. (\ref{gammashift})
is to introduce the projection
\beq
\Lambda_{{\vec m},{\vec n}}(R)~\rightarrow~(1+(-1)^{m_1+m_2})
\Lambda_{{\vec m},{\vec n}}(R)
\label{shiftedlattice}
\eeq
where $(-1)^{m_1+m_2}$ is taken inside the lattice sum. Fixing
the radii at $R=\sqrt{\alpha^\prime}$ and evaluating the sum it is
seen that (\ref{shiftedlattice}) reduces to 
\beq
\Lambda_{{\vec m},{\vec n}}(R)~\rightarrow~
{1\over{\vert\eta\vert^4}}
\left(\vert\chi_0\vert^2-\vert\chi_{1\over2}\vert^2\right)
\left(\vert\chi_0\vert^2-\vert\chi_{1\over2}\vert^2\right)
\label{projectedlattice}
\eeq
Therefore, the shift in eq. (\ref{gammashift}) does reproduce
the same number of massless states as the $\bb{Z}_2\times \bb{Z}_2$
at the free fermionic point, but the partition functions of the 
two models differ! Replacing the freely acting shift
by the shift $(A_3,A_3,A_3)$ along the momenta modes, rather than
the winding modes, induces the projection 
\beq
\Lambda_{{\vec m},{\vec n}}(R)~\rightarrow~(1+(-1)^{n_1+n_2})
\Lambda_{{\vec m},{\vec n}}(R)
\label{shiftedmomenta}
\eeq
reproducing again (\ref{projectedlattice}). 
Therefore, while each of 
the these freely acting shifts when acting on the $X_1$ manifold
does reproduce the data of the $Z_2\times Z_2$ orbifold at the free fermionic
point, none of these shifts, in fact, reproduces the $SO(12)$ lattice
which is realized at the free fermionic point.

\section{Discussion and conclusions}

Despite its innocuous appearance the connection between $X_1$ and $X_2$
by a freely acting shift in fact has profound consequences.
First we must realize that any string construction can
only offer a limited glimpse on the structure of string
vacua that possess some realistic properties. Thus, the
free fermionic formulation gave rise to three generation
models that were utilised to study issues like Cabibbo
mixing and neutrino masses. On the other hand the free fermionic
formulation is perhaps not the best suited to study
issues that are of a more geometrical character.
Now from the Standard Model data we may hypothesize
that any realistic string vacuum should possess
at least two ingredients. First, it should contain
three chiral generations, and second, it should
admit their SO(10) embedding. This SO(10) embedding
is not realized in the low energy effective field theory
limit of the string models, but is broken directly at the 
string level. The main phenomenological implication of this
embedding is that the weak-hypercharge has the canonical
GUT embedding. 

It has long been argued that the $\bb{Z}_2\times \bb{Z}_2$ orbifold naturally
gives rise to three chiral generations. The reason being that
it contains three twisted sectors and each of these sectors
produces one chiral generation. The reason that there are
exactly three twisted sectors is essentially because
we are modding out a three dimensional complex manifold, or
a six dimensional real manifold, by $\bb{Z}_2$ projections
that preserve the holomorphic three form. Thus, metaphorically
speaking, the reason being that six divided by two equals three. 

However, this argument would hold for any $\bb{Z}_2\times 
\bb{Z}_2$ orbifold
of a six dimensional compactified space, and in particular it 
holds for the $X_1$ manifold. Therefore, we can envision that
this manifold can produce, in principle, models with SO(10) gauge
symmetry, and three chiral generations from the three
twisted sectors. However, the caveat is that this manifold
is simply connected and hence the SO(10) symmetry
cannot be broken by the Hosotani-Wilson symmetry breaking
mechanism \cite{hosotani}.
The consequence of adding the freely acting shift (\ref{gammashift})
is that the new manifold $X_2$, while still admitting
three twisted sectors is not simply connected and hence
allows the breaking of the SO(10) symmetry to one of
its subgroups.

Thus, we can regard the utility of the free fermionic machinery
as singling out a specific class of compactified manifolds. 
In this context the freely acting shift has the crucial
function of connecting between the simply connected covering manifold
to the non-simply connected manifold. Precisely such a construction
has been utilised in \cite{mtheory} to construct non-perturbative
vacua of heterotic M-theory.
To use a simple analogy, we can regard the free fermionic
machinery as heavy duty binoculars, enabling us to look for minute
details on a mountain but obscuring the gross structures
of the mountain ridge. The geometrical insight on the other
hand provides such a gross overview, but is perhaps less
adequate in extracting detailed properties. 
However, as the precise point where the detailed properties
should be calculated is not yet known, one should regard
the phenomenological success of the free fermionic models
as merely highlighting a particular class of compactified spaces.
These manifolds then possess the overall structure that 
accommodates the detailed Standard Model properties. 
The precise localization of where these properties should be 
calculated, will require further understanding of the 
string dynamics. But, if the assertion that the class of
relevant manifolds has been singled out proves to be correct,
this is already an enormous advance and simplification.

In this letter we demonstrated the equivalence of the
partition function of the $\bb{Z}_2\times \bb{Z}_2$ orbifold on the
SO(12) lattice, with the model which is obtained from
eq. (\ref{zpshift}). Thus, the free fermionic 
$\bb{Z}_2\times \bb{Z}_2$ is realized for a specific form of
the freely acting shift given in eq. (\ref{gfh}).
However, as we discussed, all the models that are obtained
from $X_1$ by a freely acting $\bb{Z}_2$-shift have $(h_{11},h_{21})=(27,3)$
and hence may be connected by continuous extrapolations.
The study of these shifts in themselves may also yield 
additional information on the vacuum structure of these models
and is worthy of exploration. Such study is currently
underway and will be reported in future publications. 

\section{Acknowledgments}

I would like to heartily thank Carlo Angelantonj for collaboration
at the initial stage of this work and discussions throughout,
and Emilian Dudas and Jihad Mourad for useful discussions.
I would like to thank the Orsay theory group and LPTENS for
hospitality in the initial stage of this work. This work
is supported in part by the Royal Society and by PPARC. 

%=========================================================================
%======================== REFERENCES =====================================
%=========================================================================

\vfill\eject

\bigskip
\medskip

\bibliographystyle{unsrt}

\end{document}